\documentclass[%
 aip,
 amsmath,amssymb,
 reprint,%
]{revtex4-1}
\usepackage{hyperref}
\usepackage{graphicx}% Include figure files
\usepackage{dcolumn}% Align table columns on decimal point
\usepackage{bm}% bold math
\usepackage[utf8]{inputenc}
\usepackage[T1]{fontenc}
\usepackage{mathptmx}
\usepackage{etoolbox}
\usepackage{stackengine}
\usepackage{array}% http://ctan.org/pkg/array
\makeatletter
\def\@email#1#2{%
 \endgroup
 \patchcmd{\titleblock@produce}
  {\frontmatter@RRAPformat}
  {\frontmatter@RRAPformat{\produce@RRAP{*#1\href{mailto:#2}{#2}}}\frontmatter@RRAPformat}
  {}{}
}%
\makeatother
\usepackage{graphicx}% Include figure files
\usepackage{dcolumn}% Align table columns on decimal point
\usepackage{bm}% bold math
\usepackage{xcolor}
\usepackage{cases}
\usepackage{float}
\usepackage[overload]{empheq}

\newcommand{\hmax}{h_\text{max}}

\newcommand{\Qopt}{Q_\text{opt}}

\newcommand{\etal}{\textit{et al.}}
\newcommand{\ra}[1]{\renewcommand{\arraystretch}{#1}}
\newcommand{\micron}{\,\text{$\mu$m}}
\newcommand{\nm}{\,\text{nm}}

\begin{document}

\preprint{AIP/123-QED}
\title {Perspective on Near-Field Radiative Heat Transfer}
% Force line breaks with \\
\author{Mariano Pascale}
\author{Maxime Giteau}%
\author{Georgia T. Papadakis}
\email{georgia.papadakis@icfo.eu}
\affiliation{%
ICFO-Institut de Ciencies Fotoniques, The Barcelona Institute of Science and Technology, Castelldefels (Barcelona) 08860, Spain
}%
% \date{\today}\
\begin{abstract}
Although near-field radiative heat transfer was introduced in the 1950's, interest in the field has recently revived, as the effect promises improved performance in various applications where contactless temperature regulation in the small-scale is a requirement. With progress in computational electromagnetics as well as in nanoinstrumentation, it has become possible to simulate the effect in complex configurations and to measure it with high precision. In this Perspective, we highlight key theoretical and experimental advances in the field, and we discuss important developments in tailoring and enhancing near-field thermal emission and heat transfer. We discuss opportunities in heat-to-electricity energy conversion with thermophotovoltaic systems, as well as non-reciprocal heat transfer, as two of many recent focus topics in the field.
 Finally, we highlight key experimental challenges and opportunities with emerging materials, for probing near-field heat transfer for relevant technologies in the large-scale.
\end{abstract}

\maketitle
% \begin{quotation}
% The ``lead paragraph'' is encapsulated with the \LaTeX\ 
% \verb+quotation+ environment and is formatted as a single paragraph before the first section heading. 
% (The \verb+quotation+ environment reverts to its usual meaning after the first sectioning command.) 
% Note that numbered references are allowed in the lead paragraph.
% %
% The lead paragraph will only be found in an article being prepared for the journal \textit{Chaos}.
% \end{quotation}
\section{Introduction}\label{sec:Intro}
%\textcolor{blue}{Mitra: We need to start with "NFRHT", what you have in the 3rd paragraph, then discuss the effect of distance going through Stefan-Boltzman, earth-sun, blackbodies and then upper limits of NFRHT.}
All objects at non-zero temperature emit thermal radiation due to the thermally excited motion of particles and quasiparticles. Thermal radiation  is exchanged between objects separated by large distances, such as the earth and the sun or outer space.
 % All objects at non-zero temperature host thermally generated fluctuating currents.  These currents generate thermal radiation, which is exchanged between objects separated by large distances, such as the earth and the sun. 
 Controlling this energy exchange enables applications in solar energy harvesting, including solar photovoltaics~\cite{shockley_detailed_1961,markvart_practical_2003,wurfel_physics_2016,giteau_hot-carrier_2022} and daytime radiative cooling~\cite{li_nighttime_2021,goldstein_sub-ambient_2017,raman_passive_2014,mandal_paints_2020,zhai_scalable-manufactured_2017,zhou_polydimethylsiloxane-coated_2019}. 
% ~\cite{li_nighttime_2021,goldstein_sub-ambient_2017,raman_passive_2014,mandal_paints_2020,mandal_hierarchailly_2018,zhai_scalable-manufactured_2017,zhou_polydimethylsiloxane-coated_2019,han_preliminary_2020,wang_structural_2021}. 
 The ability to understand and control thermal radiation is key in all scenarios where radiative heat transfer (RHT) prevails over thermal conduction and convection. Examples include thermal protection systems for spacecrafts~\cite{laub_thermal_2004} and high-temperature heat exchangers~\cite{zhang_nanomicroscale_2007,howell_thermal_2020}. Tailoring thermal radiation, thus, impacts a plethora of areas of science and engineering, and research in thermal photonics is setting new frontiers for novel and disruptive technologies in the years to come ~\cite{cuevas_radiative_2018,biehs_near-field_2021}.
 
%  Understanding its underlying nature and governing laws is hence of central importance.

Besides being one of the first successes of quantum physics, Planck's law of thermal radiation ~\cite{planck_uber_1900} has been the cornerstone of modern nanophotonics at mid-infrared (IR) frequencies. The law states that a blackbody, an idealized object that perfectly absorbs radiation at all frequencies, radiates an electromagnetic spectrum which depends only on its temperature, and peaks at the thermal wavelength $\lambda_\mathrm{th}=\alpha/T$ , with $\alpha\approx 2898 \,\micron\, \text{K}$, as dictated by Wien's displacement law~\cite{planck_zur_1901}. At room temperature, $\lambda_\mathrm{th}$ is roughly $10$ microns. 

\begin{figure}[t]
    \centering
    \includegraphics[width=\columnwidth]{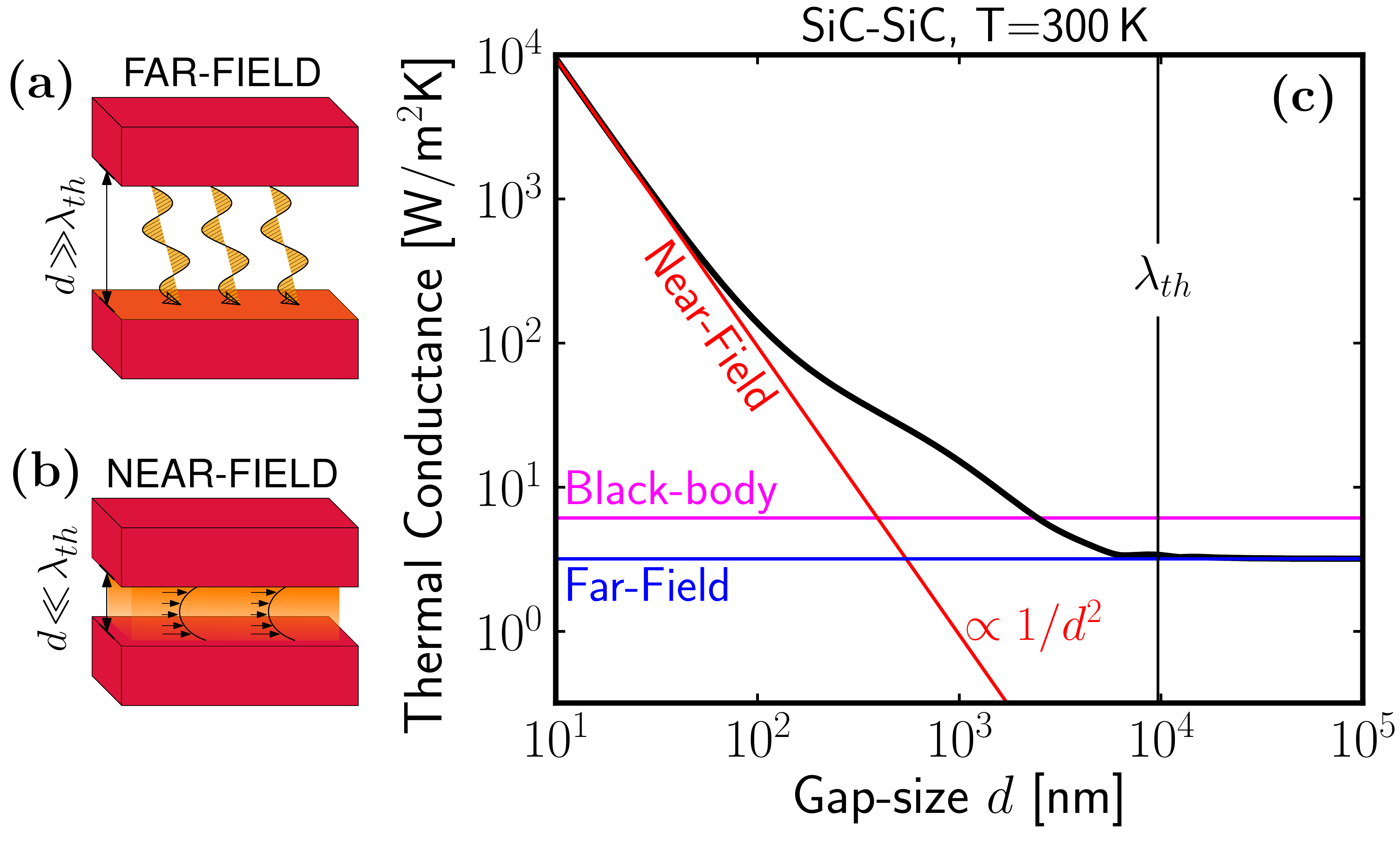}
    \caption{RHT between two bulk planar layers at different temperatures, separated by a vacuum gap of size $d$ much greater than the thermal wavelength $\lambda_\mathrm{th}$ (far-field) is mediated by propagating waves $\bf(a)$. When the two layers are brought at a distance $d\ll \lambda_\mathrm{th}$ (near-field), RHT can be mediated by evanescent modes $\bf(b)$. In panel $\bf(c)$, the radiative thermal conductance (or heat transfer coefficient, $h$) per unit area for two semi-infinite planar layers of SiC near room temperature ($T=300\,K$) is shown. For $d\ll\lambda_\mathrm{th}$, with $\lambda_\mathrm{th}\approx 10\,\micron$ marked by a vertical line, $h$ undergoes a $\propto d^{-2}$ enhancement (red curve). For $d\gg\lambda_\mathrm{th}$, $h$ saturates at its far-field value (blue line), which is smaller than the blackbody limit $\approx 6\,\text{W/m}^2 \text{K}$ (grey line).}   
    \label{fig:schematics_NF_FF}
\end{figure}

When two objects are separated by a distance $d$ much greater than $\lambda_\mathrm{th}$ (Fig. \ref{fig:schematics_NF_FF} (a)), the Stefan-Boltzmann law, a direct consequence of Planck's law, predicts that the upper bound on the net radiative heat flux is $\sigma (T_1^4-T_2^4)$, where $\sigma$ is the Stefan-Boltzmann constant, and $T_1\geq T_2$ are the temperatures of the two objects exchanging heat. This heat is exchanged via far-field propagating electromagnetic modes that oscillate harmonically as seen in Fig.~\ref{fig:schematics_NF_FF} (a).

% \textcolor{red}{KIRCHHOFF' LAW}
%  Kirchhoff's law of thermal radiation relates the emitting and absorbing capabilities of an object. Emissivity = Absorptivity.

When the heat-exchanging objects, however, are in close proximity with respect to the thermal wavelength ($d\ll \lambda_\mathrm{th}$), evanescent modes can mediate the RHT, hence Planck's law as well as the Stefan-Boltzmann law cease to hold ~\cite{basu_review_2009,song_near-field_2015,biehs_near-field_2021,polder_theory_1971,volokitin_near-field_2007}. Near-field radiative heat transfer (NFRHT) can exceed the far-field predictions by several orders of magnitude, owing to these evanescent electromagnetic modes. Such excitations are surface plasmon polaritons (SPPs) or surface phonon polaritons (SPhPs), and they carry large photon momenta~\cite{narayanaswamy_surface_2003,laroche_near-field_2006,ilic_overcoming_2012,tervo_near-field_2018,song_near-field_2015,datas_thermionic-enhanced_2019,ben-abdallah_fundamental_2010,ben-abdallah_harvesting_2019,caldwell_low-loss_2015}, termed wavenumbers henceforth. Unlike propagating modes that mediate RHT in the far-field, near-field excitations exhibit maximum intensity at the interface between two media, and decay exponentially away from it, as schematically depicted in Fig.~\ref{fig:schematics_NF_FF} (b). By placing two such interfaces in close proximity, photon tunneling occurs, explaining the large heat transfer rates predicted in the near-field as compared to the far-field \cite{polder_theory_1971,joulain_surface_2005,mulet_enhanced_2002}. 

This is shown in panel (c) of Fig. \ref{fig:schematics_NF_FF}, by plotting $h$, the thermal conductance per unit area (or heat transfer coefficient). We note that $h$ is plotted in logarithmic scale, due to the considerable difference in the heat transfer rates in the near-field as compared to the far-field. The results in Fig. \ref{fig:schematics_NF_FF} pertain to a temperature gradient near $300$ K, and the material considered in these calculations is SiC, because it supports a prominent SPhP mode near $12$ microns. As can be seen, far-field RHT is independent of the distance between the bodies exchanging heat. In contrast, NFRHT is inversely dependent on the separation distance, $d$. The upper bound of $h$ between two planar interfaces separated in the near-field can be approximated as $\displaystyle \frac{\sigma'}{d^2} (T_1^2-T_2^2)$~\cite{volokitin_resonant_2004,pendry_radiative_1999,ben-abdallah_fundamental_2010,zhang_all_2022}, where $\sigma'$ is only dependent on universal constants. As can be seen in  Fig. \ref{fig:schematics_NF_FF} (c), for gap sizes below $100$ nm, heat transfer undergoes a transition and scales as $1/d^2$ with respect to its far-field value. Recent developments in nanoinstrumentation \cite{ganjeh_platform_2012} now make it practically possible to separate objects such small distances, for which the effect has been measured various times \cite{shen_surface_2009,cui_study_2017,fiorino_giant_2018,lucchesi_radiative_2021}, as discussed below.

The scaling law describing how NFRHT depends on $d$ varies for different configurations. In Table \ref{tab:gap_dependence}, we summarize this law for some canonical configurations. Furthermore, some analytical methodologies have been developed for arbitrary shapes ~\cite{miller_fundamental_2014,venkataram_fundamental_2020}. We note that the results in Table \ref{tab:gap_dependence} are valid within a macroscopic description of thermal fluctuations. To properly model how NFRHT scales in the limit $d\to 0$, often expressed as the "extreme near-field", one ought to consider effects relevant at microscopic scales, e.g., non-local electromagnetic response as well as transport properties of the materials \cite{chapuis_effects_2008,esfarjani_heat_2011,chiloyan_transition_2015,venkataram_phonon-polariton_2018}.

The enhancement of RHT in the near-field discussed above becomes relevant in applications ranging from contactless cooling~\cite{guha_near-field_2012,kerschbaumer_contactless_2021,epstein_observation_1995} to harvesting energy with thermophotovoltaic (TPV) systems ~\cite{laroche_near-field_2006,mittapally_near-field_2021,papadakis_thermodynamics_2021},
thermal lithography~\cite{pendry_radiative_1999,howell_thermal_2020,garcia_advanced_2014,hu_tip-based_2017,malshe_tip-based_2010}, thermally-assisted magnetic recording~\cite{hamann_thermally_2004,ruigrok_disk_2000,kief_materials_2018}, thermal logic circuitry~\cite{otey_thermal_2010,fiorino_thermal_2018,ben-abdallah_near-field_2014,ordonez-miranda_radiative_2019,papadakis_gate-tunable_2019,papadakis_deep-subwavelength_2021} and scanning thermal microscopy~\cite{de_wilde_thermal_2006,kittel_near-field_2008}. In this perspective, we review experimental as well as theoretical approaches to measure NFRHT and compute the effect in complex configurations, respectively. We present our recent contributions ranging from fundamental findings such as the analytical description of NFRHT in planar structures, to more applied concepts such as actively tailoring NFRHT, opportunities of NFRHT for thermophotovoltaic systems that convert heat to electricity, and non-reciprocal heat transfer. Following, we discuss foreseen challenges for the development of NFRHT-related technologies, including achieving large-scale vacuum gaps, approaches to access spectrally and angularly-resolved information of thermal emission in the near-field,  and improving the quality of narrow-bandgap semiconductors for TPV systems. 

\begin{table}[t]\label{tab:gap_dependence}
\ra{2}
\caption{Scaling laws for NFRHT as a function of the gap-size $d$ for typical canonical geometries.}
\begin{ruledtabular}
\begin{tabular}{|c|c|c|c|}
\stackanchor{PLATE-}{PLATE}&\stackanchor{SPHERE-}{PLATE}&\stackanchor{TIP-}{PLATE}&\stackanchor{SPHERE-}{SPHERE}
\vspace{1mm}
\\
\hline
{\begin{minipage}{.1\textwidth}
\vspace{0.2mm}
      \includegraphics[width=0.8\linewidth]{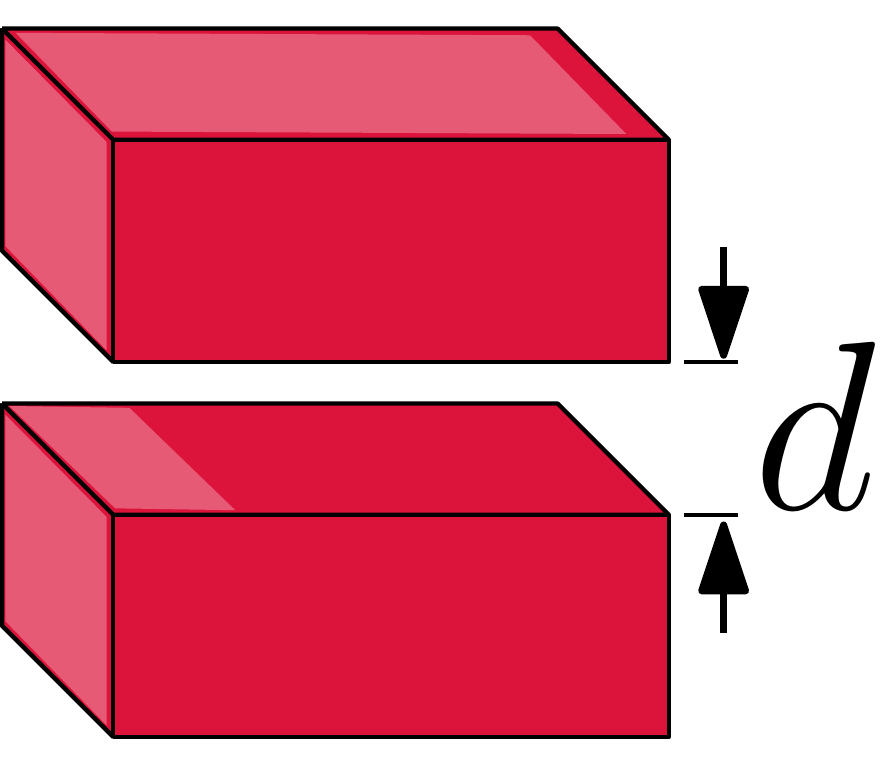}
    \end{minipage}}&{\begin{minipage}{.1\textwidth}
      \includegraphics[width=0.8\linewidth]{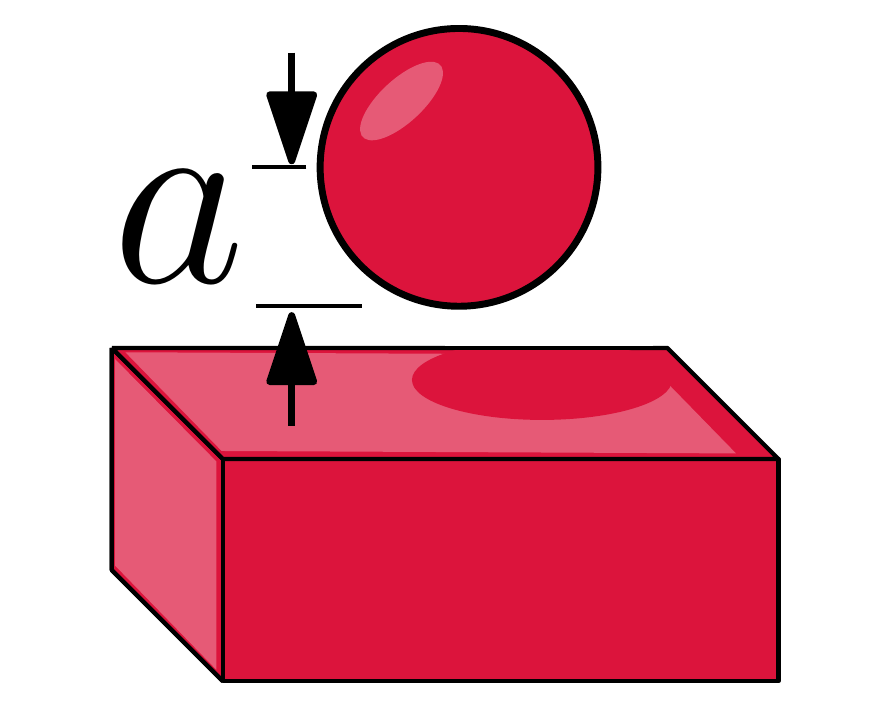}
    \end{minipage}}&{\begin{minipage}{.1\textwidth}
      \includegraphics[width=0.8\linewidth]{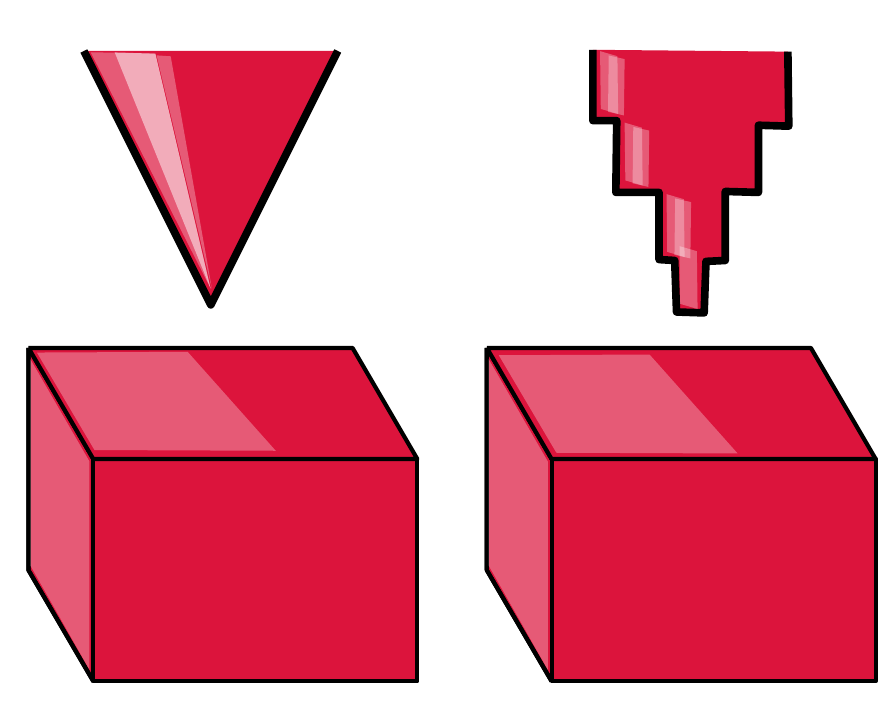}
    \end{minipage}}&{\begin{minipage}{.1\textwidth}
      \includegraphics[width=0.8\linewidth]{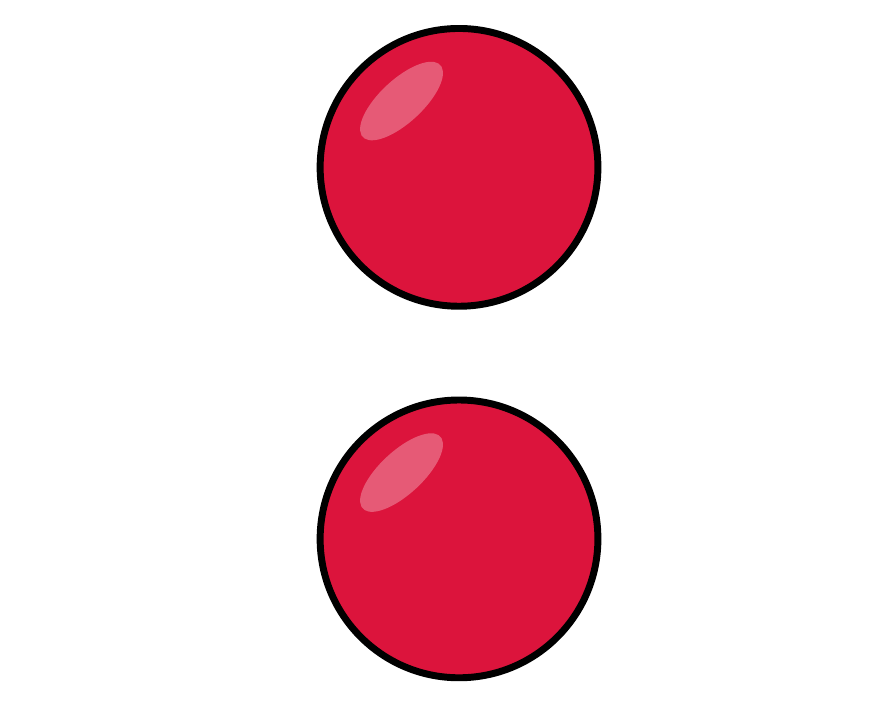}
    \end{minipage}}\\
\hline
$\displaystyle \frac{1}{d^2}$~\cite{joulain_surface_2005,fiorino_giant_2018}& $\displaystyle \frac{1}{d}\left[a\!\gg\!d\right]$\cite{golyk_small_2013,edalatpour_near-field_2016,rousseau_radiative_2009,shen_surface_2009} &$\displaystyle \log{d}$\cite{mccauley_modeling_2012}(left)&$\displaystyle \frac{1}{d}\left[a\!\gg\!d\right]$\cite{narayanaswamy_thermal_2008,chapuis_radiative_2008}\\
&$\displaystyle \frac{1}{d^3}\left[a< d\right]$\cite{mulet_nanoscale_2001}&$\displaystyle \frac{1\cite{edalatpour_near-field_2016,kloppstech_giant_2017,cui_study_2017}}{d^{\alpha\in[0.3,2]}}$(right)&$\displaystyle\frac{1}{d^{6}}\left[a\!\ll\!d\right]$\cite{narayanaswamy_thermal_2008,chapuis_radiative_2008}
\end{tabular}
\end{ruledtabular}
\end{table}

\section{Overview}

In this section, we carry out a brief review of experimental advances in measuring NFRHT, as well as theoretical approaches to model the effect and predict device performances and fundamental limits. We start by emphasizing that various research groups have recently achieved impressively small (down to few tens of nanometers) vacuum gaps, and have carried out highly precise calorimetric measurements, which has significantly contributed to the development of the field.

\subsection{Measuring RHT at the nanoscale} \label{sec:NFRHT_experiment}

NFRHT measurements are rather challenging to perform in practice. They require fine control of the separation between objects at the nanoscale ~\cite{song_near-field_2015}, as well as ensuring a uniform temperature gradient across interfaces. One of the first NFRHT experiments was performed in the late 1960s by Domoto \etal~\cite{domoto_experimental_1970}, measuring the heat exchange between two parallel copper disks at cryogenic temperatures. At such temperatures, the thermal wavelength is of the order of hundreds of microns. The authors observed RHT that surpassed the blackbody limit when the disks were placed within tens of microns from each other. The following decades have witnessed important advances in nanofabrication and nanoinstrumentation techniques \cite{ganjeh_platform_2012}. Thus, truly nanometric vacuum gaps have been achieved and the measurement of NFRHT has been possible near room-temperature. 

\begin{figure}[t]
    \centering
    \includegraphics[width=\columnwidth]{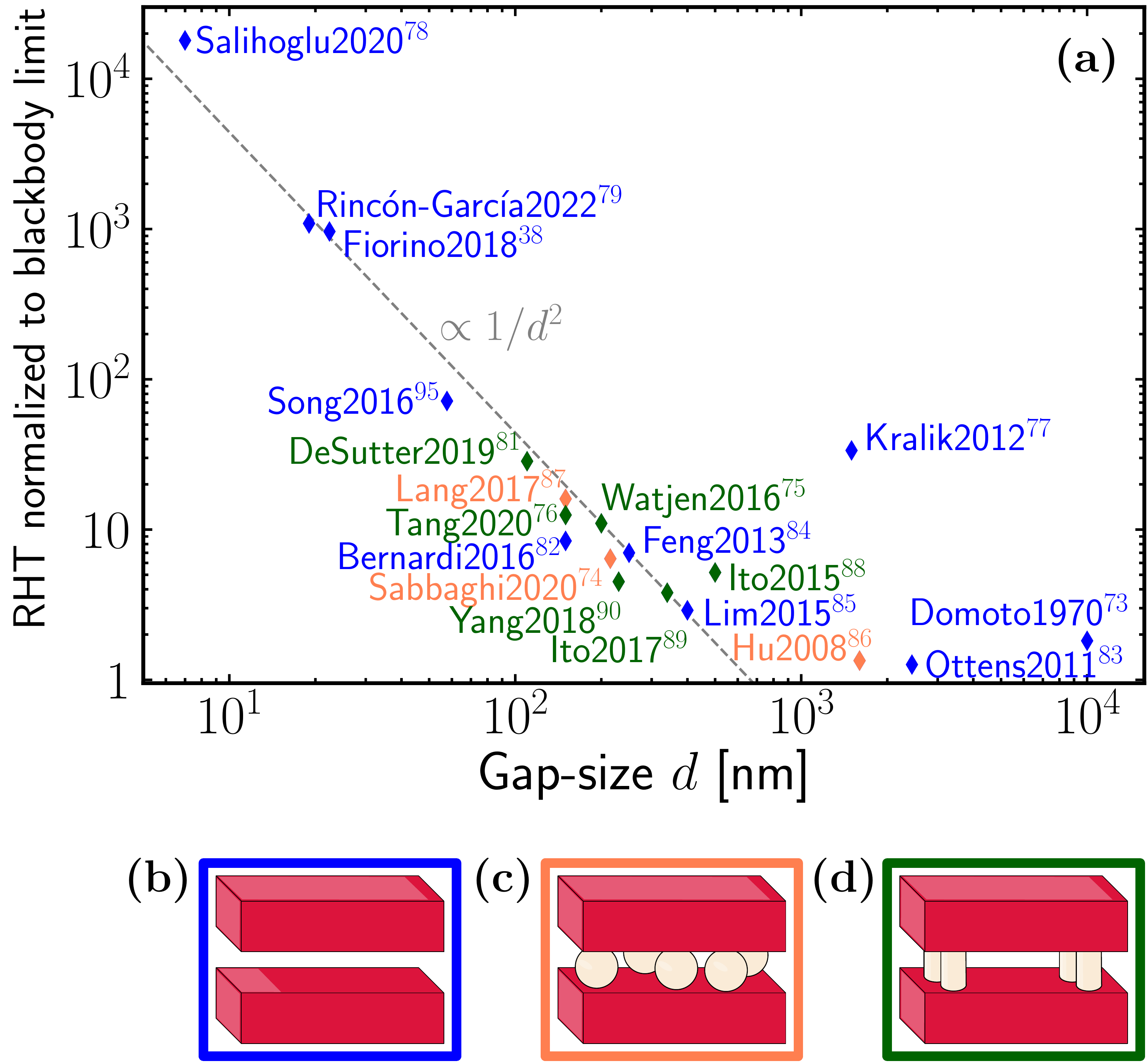}
    \caption{
  {\bf (a)} Radiative heat flux normalized to the blackbody limit as a function of the minimum gap-size $d$ for several planar configuations experimetally studied in the reported references. The vacuum gap has been achieved: {\bf (b)}  without any interposed supporting structure (blue markers)~\cite{domoto_experimental_1970,ottens_near-field_2011,kralik_strong_2012,feng_mems_2013,bernardi_radiative_2016,lim_near-field_2015,fiorino_giant_2018,salihoglu_near-field_2020,rincon-garcia_enhancement_2022}, for example using a MEMS~\cite{fiorino_giant_2018}; {\bf (c)} with dispersed nanospheres ~\cite{hu_near-field_2008,lang_dynamic_2017,sabbaghi_super-planckian_2020} (orange markers); {\bf (d)} with  micropillars~\cite{watjen_near-field_2016,ito_parallel-plate_2015,ito_dynamic_2017,yang_observing_2018,desutter_near-field_2019,tang_near-field_2020} (green markers). A qualitative $1/d^2$-trend for experiments at room temperature is also shown with a dashed line.
} 
    \label{fig:plate_to_plate}
\end{figure}

An important prototypical configuration for NFRHT, on which most theoretical studies are based, is that of two parallel plates separated by a vacuum gap (Fig.~\ref{fig:schematics_NF_FF}). Though it is arguably the simplest geometry to treat theoretically as well as a relevant one for applications, the plane-to-plane configuration is challenging to experimentally realize. This is due to the requirement of a uniform nanometric vacuum gap between smooth parallel surfaces~\cite{song_near-field_2015,ganjeh_platform_2012}. Such nanometric vacuum gaps in planar configurations have been previously achieved via suspended parallel plates, for example using micro-electro-mechanical systems (MEMS) \cite{fiorino_giant_2018}, as well as by separating the plates using nanospheres or micropillars (Fig.~\ref{fig:plate_to_plate}(b-d)
\cite{sabbaghi_super-planckian_2020,watjen_near-field_2016,tang_near-field_2020}.
% \textcolor{red}{[CITATIONS - e.g. Bo Zhao APL Doped Si experiment and others]}. 

Key NFRHT measurements reported in the literature between macroscopic plates (area $\gg \lambda_{th}^2$) are presented in Fig~\ref{fig:plate_to_plate}(a), as a function of the vacuum gap size. For any gap-dependent experiment, NFRHT follows the expected $1/d^2$ trend. Remarkably, even though the materials and configurations have varied significantly in these reports, all measurements performed at room temperature (i.e., all data points except Domoto \etal \cite{domoto_experimental_1970} and Kralik \etal \cite{kralik_strong_2012}) fall close to the same $1/d^2$ line. {Results that currently stand out are those by Salihoglu \etal~\cite{salihoglu_near-field_2020}, Rincón-García \etal~\cite{rincon-garcia_enhancement_2022}, and Fiorino \etal~\cite{fiorino_giant_2018}, who achieved a vacuum gap of merely $7\,\nm$, $19\,\nm$ and $25\,\nm$, respectively, between parallel silica surfaces, leading to up to 4 orders of magnitude enhancement over the blackbody limit.} Although still in progress, within the last decade, efforts to achieve large-scale nanometric vacuum gaps have been noteworthy, with important demonstrations so far \cite{bhatt_integrated_2020}.

Other than achieving a nanometric vacuum gap, to measure NFRHT, one ought to carry out a calorimetric measurement by applying and maintaining, throughout the measurement, a temperature gradient between two objects. In ~\cite{fiorino_giant_2018,desutter_near-field_2019}, to achieve this and measure NFRHT, the temperature of the receiver plate was maintained constant with a thermoelectric cooler while a thermoelectric heat pump was used to heat up the emitter. The receiver's and emitter's temperatures were monitored using thermistors. The heat transferred between the two plates for a given temperature difference was thus estimated from the power supplied to the heater. Alternatively, the temperature of the emitter and receiver can be maintained constant using thermoelectric heat pumps, and the heat transfer can be estimated directly from the supplied power~\cite{bernardi_radiative_2016}.

%RTH is measured by heating the emitter plate with a given heat current. The temperature reached by the emitter will depend on the heat current exchanged with the receiver and therefore give the total heat transfer.
To measure \emph{radiative} heat transfer, an additional challenge is to remove the contributions of heat convection and conduction. Suppressing convection requires the measurements to be performed in high vacuum, while suppressing conduction requires minimal physical contact between a hot emitter and a cold receiver. In all  cases, convection and conduction contributions should be subtracted from the total measured heat transfer~\cite{desutter_near-field_2019,watjen_near-field_2016}.

Some of the alignment and positioning constraints imposed by the plate-to-plate configuration can be relaxed by replacing one of the plates with a tip~\cite{williams_scanning_1986,kittel_near-field_2005} or a sphere~\cite{hu_near-field_2008,rousseau_radiative_2009,shen_surface_2009,song_enhancement_2015}. Indeed, initial modern experiments on NFRHT considered sphere-to-plane configurations \cite{narayanaswamy_near-field_2008}. Scanning thermal microscopy (SThM), which employs a tip-to-plane configuration, is one on the most notable applications of NFRHT as a measurement technique~\cite{williams_scanning_1986,wischnath_near-field_2008,worbes_enhanced_2013,kim_radiative_2015,kloppstech_giant_2017}. Akin to scanning tunneling microscopy and atomic force microscopy, SThM probes the surface of a sample through the near-field heat exchange between a heated tip and the sample's surface.
This configuration provides access to the integrated heat-flux in the extreme near-field regime, for gap-sizes below $10\nm$~\cite{worbes_enhanced_2013,kloppstech_giant_2017,kim_radiative_2015,kittel_near-field_2005}. For instance, Kittel \etal~\cite{kittel_near-field_2005} reported SThM measurements of NFRHT from surfaces of Au or GaN for tip-to-surface distances down to $\sim 1\nm$.

\subsection{Theoretical and computational methods} \label{sec:NFRHT_computations}

NFRHT is comprehensively described within the framework of fluctuational electrodynamics, as proposed by Rytov in the 1950s~\cite{rytov_theory_1959,rytov_principles_1989} and refined by Polder and Van Hove in the 1970s~\cite{polder_theory_1971}. Within this theory, thermal radiation originates from thermally excited fluctuating currents obeying the fluctuation-dissipation theorem~\cite{callen_irreversibility_1951,eckhardt_macroscopic_1984}. The latter links the currents' spatial correlation to both the dielectric properties of the medium emitting photons as well as the temperature. These microscopic fluctuating currents induce an electromagnetic field that transfers thermal energy via radiation and can be calculated via the macroscopic Maxwell's equations.

The effect of NFRHT between two- and many-body systems is, in principle, fully described within the framework of fluctuational electrodynamics ~\cite{ben-abdallah_many-body_2011, biehs_near-field_2021}. Yet, analytical solutions have been derived only in a few highly symmetric configurations, involving canonical structures such as spheres~\cite{narayanaswamy_thermal_2008,mulet_nanoscale_2001}, planes~\cite{joulain_surface_2005} and cones~\cite{mccauley_modeling_2012}. Solving the RHT problem in complex geometries is theoretically and computationally complex, requiring advanced numerical approaches. Nevertheless, by leveraging well-established computational techniques inherited from classical electromagnetic scattering theory \cite{van_bladel_electromagnetic_2007}, some computational methods have been established. These rely on spectral and finite element methods, according to the choice of delocalized or localized functions, respectively, as basis constituents for the scattering operators~\cite{biehs_near-field_2021,song_near-field_2015,forestiere_full-retarded_2018,ching_quasinormal-mode_1998,bimonte_nonequilibrium_2017,reid_fluctuation-induced_2013}. Methods include the scattering matrix approach ~\cite{bimonte_scattering_2009,messina_scattering-matrix_2011,kruger_trace_2012} as well as surface~\cite{rodriguez_fluctuating-surface-current_2013,otey_fluctuational_2014,forestiere_electromagnetic_2019} and volume~\cite{polimeridis_fluctuating_2015,jin_general_2017,forestiere_volume_2018} current formulations, including the thermal discrete-dipole approximation~\cite{edalatpour_thermal_2014,edalatpour_near-field_2016} and the finite difference time domain approach~\cite{luo_thermal_2004,rodriguez_frequency-selective_2011}.%,datas_fdtd_2013}.

Finally, recent works have carried out in-depth analyses of the upper bound limit of NFRHT in various configurations ~\cite{ben-abdallah_fundamental_2010,rousseau_asymptotic_2012,iizuka_analytical_2015,miller_fundamental_2014,venkataram_fundamental_2020,zhang_all_2022,pascale_role_2022}, demonstrating the important role of both intrinsic material properties as well as structural geometry for maximizing NFRHT.

\section{Recent contributions}

%\com{The increase in RHT provided by near-field configurations is expected to be instrumental in applications including contactless cooling~\cite{guha_near-field_2012,kerschbaumer_contactless_2021,epstein_observation_1995}, thermal lithography~\cite{pendry_radiative_1999,howell_thermal_2020,garcia_advanced_2014,hu_tip-based_2017,malshe_tip-based_2010}, thermally assisted magnetic recording~\cite{hamann_thermally_2004,ruigrok_disk_2000,kief_materials_2018}, thermal logic circuitry~\cite{otey_thermal_2010,fiorino_thermal_2018,ben-abdallah_near-field_2014,ordonez-miranda_radiative_2019,papadakis_gate-tunable_2019,papadakis_deep-subwavelength_2021}, scanning thermal microscopy~\cite{de_wilde_thermal_2006,kittel_near-field_2008} or thermophotovoltaics~\cite{laroche_near-field_2006,mittapally_near-field_2021}.}
\subsection{Analytical framework for polariton-mediated NFRHT }

In section \ref{sec:Intro}, we established that NFRHT is optimal when evanescent excitations, such as SPPs or SPhPs, are available to transfer radiative heat across a vacuum gap. Thus, plasmonic materials and polar dielectrics, respectively, are excellent candidates for near-field thermal emitters~\cite{maier_plasmonics_2007,kittel_near-field_2005,Basov2016}. 
We recently provided an analytical framework for a quantitative classification of these materials for NFRHT in the plate-to-plate configuration~\cite{pascale_role_2022}. The merit of this result lies in its simplicity: by expressing the thermal conductance as a product of three terms, $\displaystyle h = \hmax \,\Psi\left(Q\right)\Pi\left(T\right)$, the role of optical loss as well as other dielectric characteristics of a material was disentangled from the role of temperature. This offers a compact and robust description of the effect that can be used easily by the experimentalist and is in excellent agreement with fluctuational electrodynamics. 

\begin{figure}[htpb]
    \centering
    \includegraphics[width=\columnwidth]{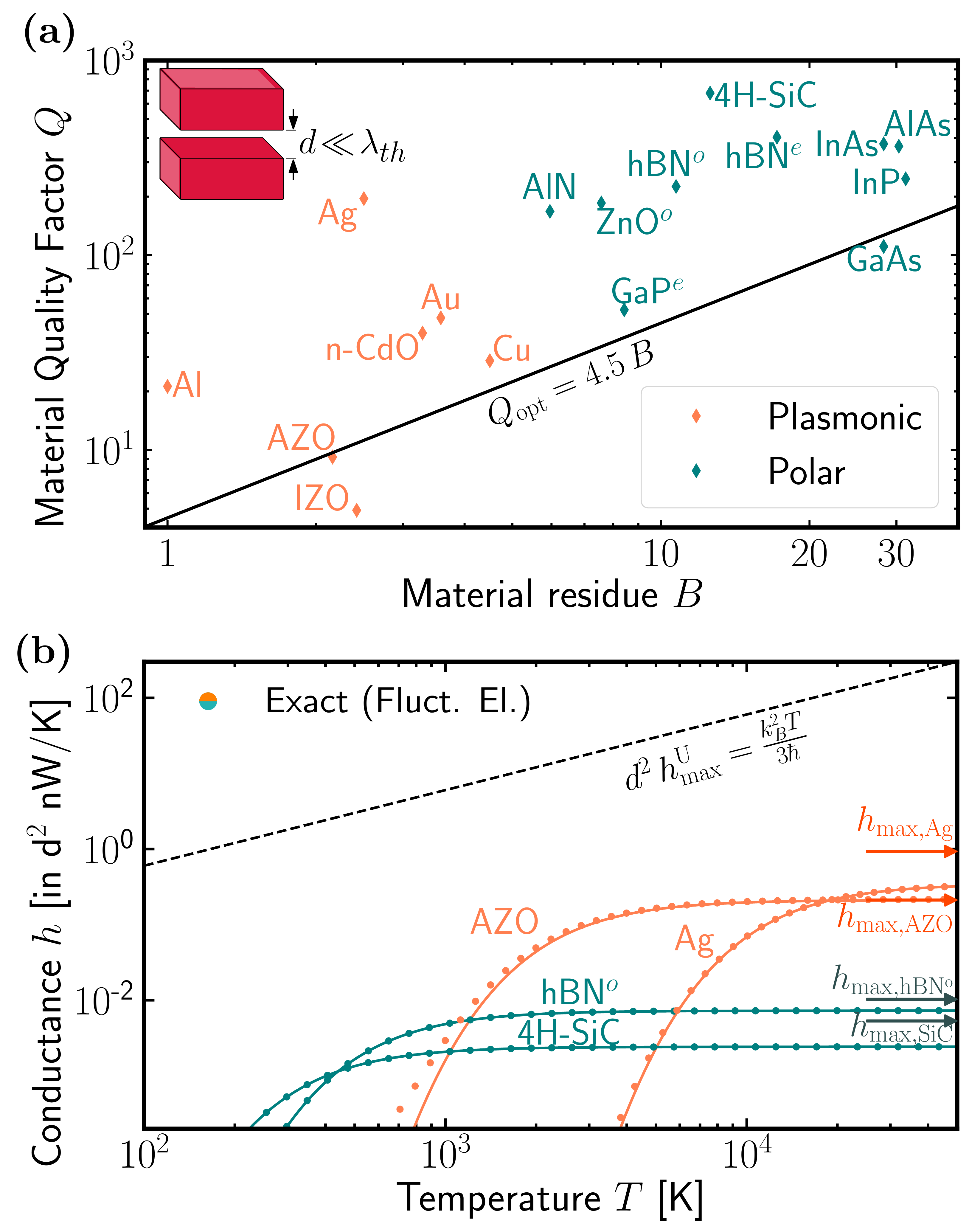}
    \caption{{\bf(a)} Quality factor $Q$~\cite{wang_general_2006,pascale_bandwidth_2021}  and material residue parameter $B$~\cite{pascale_role_2022} for plasmonic and polar materials. The Drude parameters are taken from~\cite{ashcroft_solid_1976} for Au, Ag, Cu, Al and from~\cite{caldwell_low-loss_2015} for the rest of the considered materials. Superscripts ${}^o$ and ${}^e$ stand for the ordinary and extraordinary principal axes, respectively. The solid line shows the optimal loss condition, $Q_\text{opt}=4.5\,B$, which maximizes NFRHT for a pair of closely spaced bulk planar layers (inset), independently from the temperature.
    {\bf(b)} Radiative thermal conductance, $h$, calculated using the result in~\cite{pascale_role_2022} (solid lines) and numerically, via fluctuational electrodynamics~\cite{polder_theory_1971} (dotted lines), for plasmonic (Ag, AZO) and polar (hBN$^o$, 4H-SiC) materials. The black dashed line shows the fundamental bound to $h$~\cite{ben-abdallah_fundamental_2010}. Adapted from~\cite{pascale_role_2022}.}
    \label{fig:QB_hT}
\end{figure}

The functions $\Psi$ and $\Pi$ are bounded above by unity. Hence $\hmax$, which is proportional to $\Omega/d^2$,$\Omega$ being the polariton resonance frequency, represents the maximum achievable thermal conductance, a tight upper bound. The parameter $Q$ is the material quality factor, $Q=\Omega/\gamma$ ~\cite{wang_general_2006,pascale_bandwidth_2021}, where $\gamma$ is the damping rate of the plasmon or phonon polariton. Importantly, the function $\Psi$ describes how NFRHT changes with optical loss, and is maximized at the condition $Q=\Qopt=4.5\,B$, where $B$, termed \textit{material residue}~\cite{pascale_role_2022}, is a loss-independent parameter: For plasmonic media, $B$ only depends on the high-frequency permittivity $\varepsilon_\infty$ , while for polar dielectrics it is proportional to $\Omega/\Delta\omega_{{R}}$ , where $\Delta \omega_{{R}}$ is the spectral width of the Reststrahlen band~\cite{kortum_phenomenological_1969}. Thus, in this work, one can identify the optimal plasmonic material and the optimal polar material for maximizing NFRHT.  Using this result, in Fig. \ref{fig:QB_hT}(a), we classify several relevant polaritonic media considered in the literature~\cite{cardona_fundamentals_2005,schubert_infrared_2000,caldwell_low-loss_2015,kim_optimization_2013,kim_plasmonic_2013} where $\Qopt$ against the material residue $B$.

The temperature dependence of NFRHT is described via the function $\displaystyle\Pi$, which approaches its maximum as $T\to \infty$. In contrast to Wein's displacement law in the far-field, whereby $h$ scales as $T^{3}$, in the near-field, $\Pi$ decays approximately following $\sim \left(\frac{\Omega}{T}\right)^2$. On the other hand, $\hmax$ scales with the resonance frequency $\Omega$. Therefore, plasmonic materials, which support polaritons at higher frequencies than polar dielectrics~\cite{caldwell_low-loss_2015}, can in principle reach higher NFRHT rates. However, for this to occur, they must operate at extremely high temperatures to avoid a dramatic damping in $h$. We show this in Fig. \ref{fig:QB_hT}(b) for a set of plasmonic and polar materials. 
%This can be seen in Fig. \ref{fig:QB_hT}(b), where we show $h$ computed via our compact analytical result, as well as the exact numerical result (using fluctuational electrodynamics)\cite{polder_theory_1971}, for a set of plasmonic and polar materials. 
% We also plot the fundamental limit to the thermal conductance $h$, i.e., $ \hmax^\mathrm{U} = k_B^2 T/3\hbar d^2$~\cite{ben-abdallah_fundamental_2010}, which is a loose bound compared to $\hmax$, to which $h$ actually saturates at high temperatures and for optimal loss $Q=\Qopt$.
\subsection{Active tuning of NFRHT}

The ability to actively tune heat transfer is of broad interest for any application that requires thermal control~\cite{chen_experimental_2008,wuttig_phase-change_2017,picardi_dynamic_2022}. In the near-field, a particularly relevant approach is electrostatic gate tuning~\cite{inoue_realization_2014,chen_widely_2020,duan_active_2022}. Namely, by leveraging the fact that the Debye length in doped semiconductors is comparable to the penetration depth of plasmon polariton modes in these materials, one can use electrostatic gating to achieve significant spectral shift of thermal emission spectrum in the near-field, as well as significant total heat exchange modulation between planar semiconductors. In ~\cite{papadakis_gate-tunable_2019}, we discussed this extensively, starting from a simple demonstration that NFRHT scales as $\sim \mathrm{log(\omega_\mathrm{p})}$, where $\omega_\mathrm{p}$ is the plasma frequency of the semiconductor. By applying an electrostatic bias, one can significantly change the frequency and loss of the SPP that carries NFRHT, enabling a large tuning of NFRHT. This effect has been measured experimentally in ~\cite{thomas_electronic_2019}, using graphene as a two-dimensional semiconductor. 
 
The dielectric properties of materials are often also modulated by strain, and this phenomenon is enhanced in low-dimensional materials such as hexagonal boron nitride (hBN), which is strongly strain-dependent~\cite{Lyu2019}. This strain-induced tunability yields a significant shift in the surface phonon polariton resonance of hBN. In ~\cite{papadakis_deep-subwavelength_2021}, carrying out a combination of ab initio calculations of the dielectric function of hBN and NFRHT electromagnetic calculations, we leveraged this tunability to propose a deep-subwavelength near-field thermal thermal transistor which can be switched on and off depending on the strain applied to the gate. 

Finally, we note that other active tuning strategies are compatible with NFRHT, including phase-change materials \cite{ben-abdallah_near-field_2014,van_zwol_phonon_2011} and photo-excitation \cite{kou_dynamic_2018}.

\subsection{Thermophotovoltaics}\label{sec:TPV}
A promising application for NFRHT is energy harvesting using thermophotovoltaic (TPV) systems, where the photons emitted by a hot emitter, typically at low-energies (below $1$ eV), are converted into electricity via a low-band gap photovoltaic cell ~\cite{datas_chapter_2021}. TPV is particularly relevant for waste heat recovery  as well as thermal storage~\cite{burger_present_2020, datas_advances_2022}.
It can be applied to harvest sunlight as well; solar-TPV promises efficiencies significantly higher than the Shockley-Queisser limit of conventional photovoltaics~\cite{shockley_detailed_1961, harder_theoretical_2003}. 
TPV can theoretically approach the Carnot conversion efficiency provided monochromatic thermal emission at the TPV cell bandgap, and recent advances in mid-IR spectrum engineering \cite{overvig_thermal_2021,yang_engineering_2022} can significantly contribute towards TPV systems reaching their thermodynamic limits.

However, when operating in the far field, where the emitter and the PV-cell are separated by many thermal wavelengths, TPV is strongly limited by Planck's law in terms of power density (defined as the product of current density and voltage), requiring very high emitter temperatures. By contrast, near-field TPV systems can achieve much higher power densities for the same efficiency, owing to the signficant enhancement of NFRHT with respect to blackbody radiation, as discussed above. This makes TPV systems particularly relevant for low-temperature applications~\cite{laroche_near-field_2006,tedah_thermoelectrics_2019}.

\begin{figure}
    \centering
    \includegraphics[width=\columnwidth]{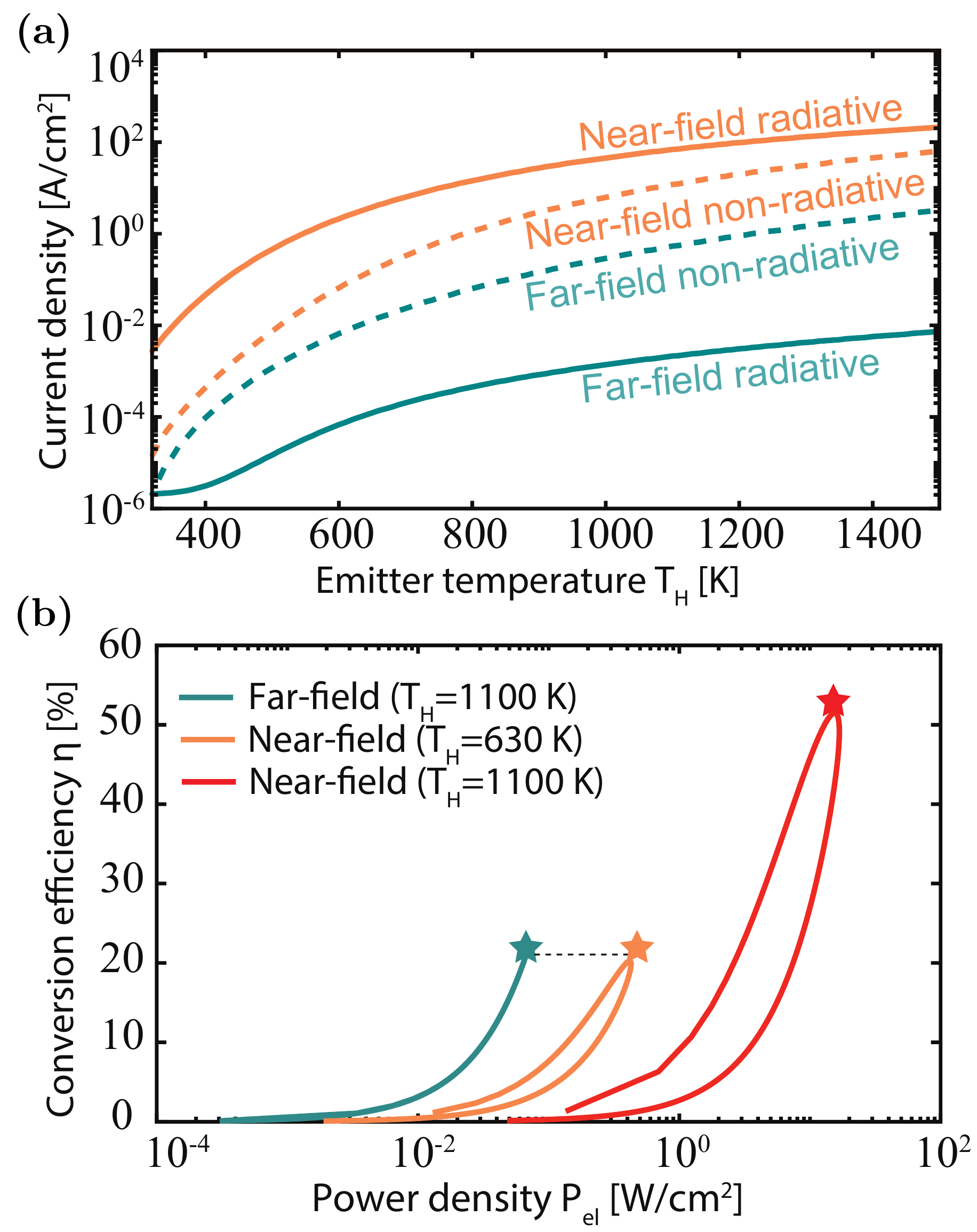}
    \caption{(a) Radiative and non-radiative recombination currents in the near-field and far-field regimes for an InAs TPV system with an ITO thermal emitter. (b) Efficiency versus power output of near-field and far-field devices. Far-field pertains to separation distances larger than 10 $\mu$m, whereas near-field pertains to a cell-emitter separation of $d=10$ nm. Asterisks pertain to the optimal operation point. Adapted from Ref.~\cite{papadakis_thermodynamics_2021}.}
    \label{fig:Papadakis}
\end{figure}

In a previous work~\cite{papadakis_thermodynamics_2021}, we showed that near-field TPV systems have some attractive advantages as compared to standard far-field operation. In particular, the carrier density achieved in near-field FTPV systems is much larger than in far-field devices. Although this leads to significant non-radiative losses, near-field-enhanced radiative currents dominate as shown in Fig.~\ref{fig:Papadakis} (a), thus yielding a near-unity luminescence efficiency. By additionally proposing thin cells in near-field TPV systems as compared to $\mu$m-thick cells used in standard solar PV and TPV, we showed that near-field TPV systems can achieve much higher power densities for the same conversion efficiency as far-field TPV ones, while requiring much lower emitter temperatures (Fig.~\ref{fig:Papadakis} (b)). In another study~\cite{papadakis_broadening_2020}, by carrying out simple detailed balance analytical calculations, we demonstrated that broadening the thermal emission spectrum (e.g. by combining several resonant emitters) can enhance both the efficiency and the power density when accounting for non-radiative recombinations. This is in contrast to the common assumption that a monochromatic spectrum will yield maximum PV conversion efficiency, an argument that holds only in the radiative limit.

\subsection{Non-reciprocity}\label{sec:nonreciprocal}
Conventional thermal emitters are made of materials satisfying Lorentz reciprocity, characterized by symmetric dielectric permittivity and magnetic permeability tensors. These reciprocal emitters have identical absorptivity and emissivity for any frequency, direction and polarization, as dictated by Kirchhoff's law for thermal radiation~\cite{kirchhoff_ueber_1860}. Kirchhoff's law can be violated by considering non-reciprocal thermal emitters, for instance made of gyrotropic materials~\cite{miller_universal_2017}. In this case, the symmetry between thermal emission and absorption is broken, leading to interesting physics and applications.

For instance, non-reciprocity is key for solar energy harvesting~\cite{fan_thermal_2017,park_reaching_2022}, where it is required to approach the so-called Landsberg conversion efficiency limit ~\cite{landsberg_thermodynamic_1980} of $93.3\%$. Breaking Kirchhoff's law has been a long-sought milestone in nanophotonics, with important recent developments \cite{shayegan_nonreciprocal_2022}. Non-reciprocity is also considered for efficient optical isolators~\cite{Asadchy_sub-wavelength_2020}. 

In ~\cite{fan_nonreciprocal_2020}, we investigated the implications of non-reciprocal thermal emission for NFRHT. We showed that, indeed, gyrotropic materials with evanescent modes can lead to an asymmetric NFRHT per frequency and wavenumber. Nevertheless, the total, integrated heat flow between two bodies at equilibrium must be the same due to thermodynamic constrains~\cite{fan_nonreciprocal_2020}. In contrast, non-reciprocity becomes relevant when at least three bodies are involved, in which case a persistent heat current can be sustained~\cite{zhu_persistent_2016,zhu_theory_2018,biehs_near-field_2021}, which has been applied to demonstrate near-field thermal hall effect~\cite{ben-abdallah_photon_2016}. 

\section{Challenges and opportunities}
 In this section, we outline challenges and opportunities in NFRHT, with the objective of optimally leveraging the latest theoretical and experimental developments to bring NFRHT closer to real-world applications. We start by what we view as the most critical aspect of NFRHT experiments, which is ensuring a large-scale, mechanically robust vacuum gap for precision measurements. Next, we discuss approaches for conducting spectroscopy measurements in the near field, especially for plane-to-plane configurations. We also highlight important emerging material and relevant properties for directional and enhanced NFRHT. Finally, with low-temperature waste heat recovery in mind, we highlight the importance of searching for high-quality low-band gap semiconductors for near-field TPV systems.

\subsection{Large-scale vacuum gap}\label{sec:vacuum_gap}

For all NFRHT applications, the ability to realize and maintain a nanometric vacuum gap is critical. This is particularly challenging in plate-to-plate configurations which require flat and parallel surfaces. Furthermore, a high vacuum (typically in the order of $10^{-6}$ mbar) is required to ensure negligible convection through air molecules or dust. Finally, parasitic conductive heat transfer channels should also be suppressed as much as possible.

So far, the vacuum gap has been achieved in various ways (see Fig.~\ref{fig:plate_to_plate}). For instance, micro-mecanical, suspended platforms have been reported in many works~\cite{ganjeh_platform_2012,st-gelais_demonstration_2014,lim_near-field_2015,st-gelais_near-field_2016,bernardi_radiative_2016,zhu_near-field_2019,mittapally_near-field_2021} (Fig.~\ref{fig:plate_to_plate}(b)). This approach is ideal for experiments as it allows proving NFRHT in the absence of heat conduction. Furthermore, the width of the vacuum gap can be varied down to less than 100 nm~\cite{fiorino_giant_2018}.
However, such systems are currently restricted to small areas with limited scalability, with a maximum reported surface area between two plates in the order of $10^4 \ \mathrm{\mu m^2}$~\cite{lucchesi_radiative_2021}. One way to achieve much larger areas is by combining many devices in parallel, as was proposed with the fully integrated architecture using a suspended-emitter bridge developed by Bhatt \textit{et al.}~\cite{bhatt_integrated_2020}.

Polystyrene or silica micro-spheres embedded in between parallel plates have also been employed as spacers to create vacuum gaps of a few hundreds of nanometers~\cite{hu_near-field_2008,lang_dynamic_2017,sabbaghi_super-planckian_2020} (Fig.~\ref{fig:plate_to_plate}(c)). The contact area of each sphere is small, hence limiting the parasitic heat conduction. It is however difficult to control their dispersion characteristics, and they also introduce parasitic radiative transmission channels.

Finally, micro-pillars~\cite{dimatteo_microngap_2004,ito_parallel-plate_2015,watjen_near-field_2016,ito_dynamic_2017,yang_observing_2018,desutter_near-field_2019,tang_near-field_2020} or trenches~\cite{inoue_integrated_2021} have been used as a spacer between parallel plates to achieve a controlled vacuum gap in a simple and scalable way (Fig.~\ref{fig:plate_to_plate}(d)). The main drawback of this approach is that it leads to heat conduction channels between the emitter and the receiver. The pillars therefore need to be sparse (often leading to significant bowing) and have the smallest heat conductivity possible. DeSutter \textit{et al.} have employed an ingenious configuration to reduce conduction. By etching pits into the emitter, the pillars can be made much longer, significantly reducing their contribution to heat transfer from 45\% down to 1.9\%~\cite{desutter_near-field_2019}. 

Given the increasing number of experimental groups working on NFRHT and owing to progress in nanomechanics, we foresee that the challenges faced in previous years in realizing nanometer-scale vacuum gaps will be overcome. This will allow access to vacuum gaps on the order of few nanometers ($1-10$ nm), also termed "extreme" near-field \cite{kim_radiative_2015}, where experimental measurements differ from the predictions of fluctuational electrodynamics~\cite{kittel_near-field_2005}. This is particularly exciting to study, as in this region it is yet unclear whether thermal conduction or radiation dominates, which is also dependent on the electronic structure of the participating materials~\cite{chiloyan_transition_2015}.

\subsection{Near-field spectrally resolved thermal spectroscopy}

As mentioned in Sec. \ref{sec:NFRHT_experiment}, most works that probe NFRHT do so using calorimetry. In other words, they measure the total heat flux between an emitter and a receiver, thereby integrating spectral and angular information. Given the strong narrowband contributions from polaritons that mediate NFRHT, the ability to spectrally resolve NFRHT is very valuable. This requires, however, to couple the evanescent waves exchanged between a hot emitter and a cold receiver to the far field so they can be measured by a standard spectrometer.

So far, this has been achieved mainly using a sharp tip as the cold receiver as well as the scatterer that couples near-field radiation to the far field. This is termed \emph{thermal radiation scanning tunneling microscopy}~\cite{de_wilde_thermal_2006,jones_thermal_2012,jones_thermal_2013,joulain_strong_2014}. However, in this tip-to-plane configuration, it is difficult to assess the exact impact of the tip on heat transfer due to the complex shape and potential irregularities. In order to tailor plate-to-plate radiative heat transfer, one work has taken an alternative approach~\cite{zare_measurement_2019}, coupling the evanescent waves from an emitter to the far-field through a high-index element with a $45^{\circ}$ bevel, in a configuration analogous to Otto coupling to surface modes~\cite{maier_plasmonics_2007}. A similar approach was considered to enhance thermal emission using a high-index transparent hemisphere~\cite{yu_enhancing_2013}. 

Furthermore, NFRHT is dominated by evanescent modes with very large wavenumbers, that extend far beyond the light line. Therefore, fully characterizing the effect of NFRHT requires not only near-field spectroscopy (resolving the frequency dependence of thermally excited modes), but also probing the effect as a function of the wavenumber. We foresee such measurements to be possible in the near future.

\subsection{NFTPV: narrow-bandgap semiconductors}
We now comment on what we view as one of the key bottlenecks in TPV development, since TPV systems are an important potential application of NFRHT as discussed in Section \ref{sec:TPV}.

In the far-field, a recent $40\%$ efficiency record~\cite{lapotin_thermophotovoltaic_2022} has confirmed that TPV can compete with other heat-to-electricity conversion technologies~\cite{datas_thermophotovoltaic_2017}.
In the near-field, however, only a handful of studies have reported TPV devices so far~\cite{inoue_one-chip_2019,fiorino_nanogap_2018,bhatt_integrated_2020,lucchesi_near-field_2021,mittapally_near-field_2021}. 
{
For instance, in \cite{fiorino_nanogap_2018} the authors experimentally studied NFTPV devices characterized by
a $\sim 40-$fold enhancement in the power output at nominally 60 nm gaps relative to the far field. 
In \cite{mittapally_near-field_2021}, record power densities of $\sim 5$ 
$\mathrm{kW/m^2}$ at an efficiency of $6.8\%$ are experimentally demonstrated, where the efficiency of the system is defined as the ratio of the electrical power output of the PV cell to the radiative heat transfer from the emitter to the PV cell.
Interestingly, in \cite{inoue_integrated_2021}, comparisons between the simulations and experiments reveal the possibility of a system efficiency of $>35\%$ in up-scaled devices, thus demonstrating the potential of NFTPV device for practical use in the future.
}

One of the major constraints for NFTPV systems is that, to prevent parasitically heating the PV-cell or driving its nonradiative losses too high, the emitter temperature ought to remain below $\mathrm{700 ^\circ C}$, otherwise Joule heating can even burn the PV-cell itself. As a result, the PV absorber must be made of a very narrow-bandgap semiconductor (below $0.5$ eV) to harvest the low-energy radiation. Unfortunately, however, narrow-bandgap semiconductors have significant nonradiative losses, thus typically yielding lower efficiency~\cite{yang_narrow_2022}. For example, although Lucchesi \textit{et al.} achieved 14\% conversion efficiency with an InSb cell (bandgap of $0.23$ eV), the cell temperature had to be maintained at 77 K to reduce non-radiative recombinations~\cite{lucchesi_near-field_2021}.

This demonstrates the need for developing materials with improved luminescence efficiently operating in the mid-IR, for sufficient absorption and efficient conversion of heat into electricity. Such narrow-bandgap semiconductors have so far received far less attention than materials like Si or GaAs, therefore material quality can still be significantly improved. Along with the fact that cells tend to operate closer to the radiative limit in the near-field~\cite{papadakis_thermodynamics_2021}, we believe that room-temperature devices with reasonable efficiencies and power densities should be achievable in the near future. 

{\subsection{Controlling NFRHT through photon chemical potential}
In a semiconductor under external bias, the photons with energy above the bandgap carry a nonzero chemical potential depending on the applied voltage. Thus, the spectral power density of radiation emitted by a diode can be enhanced (forward bias) or suppressed (reverse bias), enabling the active modulation of RHT \cite{zhao_chemical_2020}. 

Such concept has been proposed to boost the performance of cooling devices. In fact, a forward biased light emitting diode (LED) can be used for cooling purposes by radiating to the far-field part of the energy taken from the surroundings in the form of heat \cite{tauc_share_1957}. Unfortunately, the intrinsic nonidealities of far-field configurations, such as non-radiative recombination and difficulty in photon extraction, have so far hindered an experimental demonstration. Such complications can be overcome in a near-field configuration \cite{chen_heat-flux_2015,chen_high-performance_2017,song_near-field_2020,liu_high-performance_2016,xiao_electroluminescent_2018,yang_near-field_2022}. For instance, in \cite{chen_high-performance_2017}, a device consisting of a GaAs LED and a Si PV-cell at close proximity has been envisioned to deliver a cooling power density up to $10^5\,\mathrm{W/m}^2$ even in the presence of realistic Auger recombination and Shockley-Read-Hall recombination. Moreover, building on previous theoretical works \cite{chen_near-field_2016,chen_heat-flux_2015,chen_high-performance_2017}, in \cite{zhu_near-field_2019} a structure including an InAsSb LED and a Si PV-cell at nanometric distance has experimentally proved active cooling by tuning the chemical potential of photons.

The chemical potential of photons can be also exploited for power generation \cite{green_third_2001,zhao_chemical_2020}. In fact, self-sustaining systems with multiple diodes undergoing radiative exchange among them, can yield high power density for power harvesters. Among these schemes, there has been great interest toward thermophotonic (TPX) devices \cite{harder_thermophotonics_2003}. As opposed to TPV devices, which utilize a passive thermal emitter on the hot side, TPX ones use an LED, whose electro-luminescence leads to a higher radiation intensity.
It can be shown that, for a sufficiently high external electro-luminescence quantum efficiency, the increase in the power density by using an LED is much stronger than the power consumption needed to bias the LED itself \cite{harder_thermophotonics_2003}. 
When operating in the near-field regime, TPX systems (or NF-TPX) can undergo a boost in the radiative heat transfer due to tunneling waves, enabling a dramatic increase in the electrical power output. As anticipated in Sec. \ref{sec:TPV}, the near-field configuration is advantageous with respect to the far-field one in terms of efficiency as well as power density, since radiative recombination dominates (see Fig. \ref{fig:Papadakis}) \cite{papadakis_thermodynamics_2021}. Therefore, it can be shown that the thermophotonic idea can prove ideal for low-grade heat recovery applications \cite{zhao_near-field_2018,legendre_gaas-based_2022}.}
\section{Conclusions}

Near-field radiative heat transfer is an active field of research with relevance in both fundamental and applied physics. Although it was originally investigated as a means to overcome limitations in radiative power transfer, it is now also used to explore unique regimes not accessible in the far-field.
In this Perspective, we highlighted key theoretical and experimental achievements of the past years. We have presented recent contributions in the theoretical modelling of NFRHT and its applications, active tuning of NFRHT using emerging materials, detailed balance analysis of thermophotovoltaic systems, and non-reciprocal heat transfer. We further identified key challenges in harnessing NFRHT for energy conversion, paying particular attention to the progress made in achieving nanometric vacuum gaps, as well as highlighting the importance of further improving these efforts for large-scale integration. We emphasized the need to complement conventional heat transfer measurements with spectrally resolved near-field characterization techniques, which could enable a direct comparison between theory and experiments, as well as and a deeper understanding of the underlying physics.
We believe that NFRHT experiments will soon bring about the promises of theoretical predictions, leading to disruptive contactless heat transfer technologies.

\section*{Acknowledgments}
The authors thank Dr. Mitradeep Sarkar for valuable discussions. The authors declare no competing financial interest. G. T. P. acknowledges funding from ”la Caixa” Foundation (ID 100010434), from the PID2021-125441OA-I00 project funded by MCIN /AEI /10.13039/501100011033 / FEDER, UE, and from the European Union’s Horizon 2020 research and innovation programme under the Marie Sklodowska-Curie grant agreement No 847648. The fellowship code is LCF/BQ/PI21/11830019. M.G. acknowledges financial support from the Severo Ochoa Excellence Fellowship.
This work is part of the R\&D project CEX2019-000910-S, funded by MCIN/ AEI/10.13039/501100011033/ , from Fundació Cellex, Fundació Mir-Puig, and from Generalitat de Catalunya through the CERCA program.

\section*{Bibliography}
\bibliography{references_final.bib}% Produces the bibliography via BibTeX.

\end{document}